\documentclass{article}
\usepackage{srcltx}
\usepackage{amsmath}
\usepackage{amssymb}
\usepackage{amsfonts}
\usepackage{latexsym}
\usepackage{textcomp}
\usepackage{appendix}
\usepackage{multirow}
\usepackage{booktabs}
\usepackage{url}
\usepackage{array}
\usepackage{marvosym}
\usepackage{graphics}
\usepackage{graphicx}
\usepackage{subfigure}
\usepackage{epsfig}
\newcommand{\N}{{\mathbb N}}
\usepackage{color}
\usepackage{natbib}
\newtheorem{hypo}{Hypothesis}
\newtheorem{def0}{Definition}

\title{Spurious seasonality detection: a non-parametric test proposal}

\author{Aurelio F. Bariviera \\
\footnotesize Department of Business, Universitat Rovira i Virgili, Av. Universitat 1, 43204 Reus (Spain) \\ \footnotesize \texttt{aurelio.fernandez@urv.cat} \\
 Angelo Plastino \\
 \footnotesize IFLP-CONICET-UNLP, C. C. 727, 1900 La Plata (Argentina) \\
\footnotesize SThAR - EPFL Innovation Park, Lausanne (Switzerland)  \\
 George Judge \\
\footnotesize Professor of the Graduate College, 207 Giannini Hall, University of California Berkeley (United States)
}

\begin{document}

\maketitle

\begin{abstract}
This paper offers a general and comprehensive definition of the day-of-the-week effect.
 Using symbolic dynamics, we develop a unique test based on ordinal patterns in order to detect it.
 This test uncovers the fact that the so-called ``day-of-the-week'' effect
  is partly an artifact of the hidden correlation structure of the data.  We present simulations based on artificial time series as well.
  Whereas time series generated with long memory are prone to exhibit daily seasonality,
  pure white noise signals exhibit no pattern preference. Since ours is a non parametric test, it requires no assumptions about the
  distribution of returns so that it could be a practical alternative to conventional econometric tests.
   We made also an exhaustive application of the here proposed technique to 83 stock indices
    around the world. Finally, the paper highlights the relevance of symbolic analysis
    in economic time series studies. 
\end{abstract}%

\section{Introduction \label{sec:intro}}
The static capital asset pricing model (CAPM), developed
independently by \cite{Sharpe64}, \cite{Lintner65} and
\cite{Mossin66}, has been widely used for a number of financial
matters. In its standard form, the CAPM states that the expected
risk on a security $i$ can be separated into two components: the
risk free rate and the risk premium. The latter, in turn, can be
explained as the product between the market premium and a modulating
coefficient $\beta$:
\begin{equation}
E(R_i)=R_f+\beta_i \left( E[R_m]-R_f \right)
\label{eq:capm}
\end{equation}
According to equation \ref{eq:capm}, return on the security $i$
depends only on the risk free rate, market return, and beta.
Consequently, return should not be altered by any other
circumstances such as the particular day of the week or the time of
the year in which the return is measured. According to \cite{Fama70}
a market is informationally efficient if it fully reflects all
available information. In fact, as \cite{LeRoy} asserts, the
Efficient Market Hypothesis (EMH) is just the idea of competitive
equilibrium applied to the securities market.

Although early empirical studies (e.g.
\cite{BlumeFriend73,FamaMacBeth73}) support  the validity of the
CAPM, later research documents departures from this equilibrium
model. These departures are called ``anomalies'' \footnote{According
to \cite{Kuhn68}, an anomaly is a fact that puts into question an
established paradigm.}. Among them, there is one specially puzzling
feature, the day-of-the-week effect. This anomaly refers to the
heterogeneous behavior of returns along the week. Testing the the
day-of-the-week effect requires the joint consideration of an
equilibrium model, such as equation \ref{eq:capm}, and of the
Efficient Market Hypothesis (EMH).

Empirical research on markets' daily seasonality can be traced back
to \cite{Fields31,Fields34}. These papers have the merit of
investigating the issue before a market equilibrium model was
formally developed. \cite{Cross73} detects differences in expected
S\&P 500 on Fridays and Mondays. \cite{GibbonsHess81} finds lower
S\&P 500 returns on Mondays relative to other days. The effect is
subdivided by \cite{French80} into a Monday effect (abnormal
negative return on this day) and a Friday effect (abnormal positive
result on this day). \cite{Rogalski84} analyzes the effect during
trading and non-trading hours for the American market. This effect
has been widely surveyed and, for brevity, we refer to
\cite{ZiembaKeim00} and \cite{Ziemba2012} for further discussion on
empirical works about this effect. \cite{KELOHARJU2016} find return seasonalities in commodities and stock indices arround the world.

The standard approach for detecting the day-of-the week effect is
based on  (or some variations of) the following regression equation:
\begin{equation}
r_t=\alpha_0 +\alpha_1 D_{1t}+\alpha_2 D_{2t}+\alpha_3 D_{3t}+\alpha_4 D_{4t}+\epsilon_t
\label{eq:dow}
\end{equation}
where $r_t$ is the return on day $t$ and $D_i$, $i=\{1\dots,4\}$ are
dichotomous dummy variables for each day of the week from Tuesday
through Friday. The coefficient $\alpha_0$ represents the mean
return on Monday, while $\alpha_i$,  $i=\{1\dots,4\}$, is the excess
return on  day $i$, and $\epsilon_t$ is an error term. This
traditional  approach is based on different hypothesis testing on
$\alpha_i$ values (for an overview see, for instance,
\cite{Bariviera05} and references therein).  Working based on Equation \ref{eq:dow}
forces to make several (sometimes unjustified) assumptions about parameters. For example, \cite{ZHANG2017} applies a rolling sample test with a GARCH model in 28 stock indices. Precisely, our original approach, based on ordinal patterns, bypasses this shortcome.  

The aim of this paper is  to provide a more general definition of
the day-of-the-week effect and to develop an alternative test to
assess the existence of seasonal effects in daily returns. This
paper contributes to the literature in several ways. First, it
generalizes the definition of the day-of-the-week effect. Second, it
develops an alternative non-parametric test to detect it. Third, it
shows that the results of the tests are not obtained by chance,
since time-causality is taken into consideration. Fourth, most of
the day-of-the-week findings in the literature are related with the
underlying return-generating process, and not with the causes
indicated previously in the literature. Consequently, from a theoretical point of view, this paper introduces a new non-parametric test that is able to detect the intrinsic characteristics of the time series, and uncover spurious seasonality detection causes.
We would like to point out that the methodology we use here is unique in that it is nonlinear, ordinal, requires no model 
and provides statistical results in term of a probability density function. Ours is a statistical methodology
that, to the extent of our knowledge, no one using time series analysis has used it before.

The remaining of the paper  is organized as follows. Section
\ref{sec:op} presents the notion of ordinal patterns. Section
\ref{sec:dow} redefines the day-of the week effect and proposes a
non-parametric test. Section \ref{sec:simulation} displays results
of the test on theoretical simulations of different stochastic
processes. Section \ref{sec:empirical} performs an empirical
application to the New York Stock Exchange. Finally, Section some
conclusions are drawn in \ref{sec:conclusion}.

\section{Ordinal pattern analysis \label{sec:op}}
Estimations based on equation \ref{eq:dow}  require the assumption
of an underlying stochastic process for returns. For these
processes, symbolic analysis becomes a suitable alternative to study
the dynamics of a time series. \cite{BandtPompe02} develop a method
for estimating the probability  distribution function (PDF) based on
counting ordinal patterns. The comparison of neighboring values of a
time series requires no model assumption \cite{BandtPompe93}. The
advantage of this method is that can be applied to any time series,
and takes into account time causality \cite{BandtPompe93}. If returns fulfill the Efficient Market Hypothesis (EMH),  there should be
no privileged pattern. If there were, it would be exploited by
arbitrageurs and any possibility of abnormal return should be
rapidly wiped out. Thus, if the time series is random, patterns'
frequency should be the same, provided $N \gg D!$

If patterns are not equally present in the sample,  three anomalous
situations might be  the cause:
\begin{itemize}
\item[i.] Forbidden pattern: a pattern that does not appear within the sample.
\item[ii.] Rare pattern: a pattern that seldom appears.
\item[iii.] Preferred pattern: a pattern that emerges  oftener  than expected by the uniform distribution.
\end{itemize}
In any of such cases we are  in presence of a time series with daily
seasonal behavior. Consequently the day-of-the week effect needs to
be redefined.

In this line, \cite{Zanin08} applies the concept of forbidden patterns in
order to assess market efficiency, and shows that different
financial instruments could achieve different informational
efficiency. According to \cite{Amigoetal2006}, forbidden patterns
can be used as a means of distinguishing chaotic and random
trajectories and constitute a satisfactory alternative to more
conventional techniques.

Ordinal patterns have been previously used by
\cite{ZuninoCausality10,ZuninoPermutation11,Zunino2012} in order to
compute quantifiers like permutation entropy and permutation
complexity, which, in turn, allow one to quantify the degree of
informational efficiency of different markets. \cite{RossoCarpiB}
demonstrates that forbidden patterns are a deterministic feature of
nonlinear systems. \cite{Bariviera11,BaGuMa2012} show that the
correlation structure and informational efficiency are not constant
through time and could be affected by several factors such as
liquidity or economic shocks.

Given a time series of daily returns \footnote{Let us assume that
the time series is characterized by  a continuous distribution.}
beginning on Monday ${\mathcal R}(t) = \{ r_t ; t = 1, \cdots , N
\}$ and a pattern length $D =5$, following the \cite{BandtPompe02}
method, $N/5$ partitions of the time series could be generated. Each
partition is a 5-dimensional vector $\left( r_t, r_{t+1},  r_{t+2},
r_{t+3}, r_{t+4}\right)$, which represents a whole trading week.
Each return is associated with a day of the week. For simplicity, we
have $day=\{i; i=0,  \dots, 4; i \in \N \}$ standing for Monday
through Friday. The method sets the elements of each vector in
increasing order. Doing so, each vector of returns is converted into
a symbol. For example, if in a given week
$r_{Mo}<r_{Fr}<r_{Tu}<r_{Th}<r_{We}$, where $r_{Xx}$ represents
return on day $Xx$, the pattern is $(0,4,1,3,2)$. There are $5!=120$
possible permutations. Each permutation produce a different pattern
($P$) and the associated frequencies can be easily computed. Each
pattern has a frequency of appearance in the time series.
\cite{Carpi20102020} asserts that in correlated stochastic
processes, pattern-frequency observations do not depend only on the
time series' length but also on the underlying correlation
structure. \cite{AmigoZambranoSanuan07,AmigoZambranoSanuan08} show
that in uncorrelated stochastic processes every ordinal pattern has
an equal probability of appearance.
                                                                                                                                             Given that the ordinal-pattern's associated PDF is invariant with respect to nonlinear monotonous transformations, \cite{BandtPompe02} method results suitable for experimental data (see e.g. \cite{Saco2010,Parlitz2012}). A graphical meaning of the ordinal pattern can be seen in \cite{Parlitz2012}.

\section{Day-of-the-week effect: a redefinition of the problem \label{sec:dow}}

As recalled in Section \ref{sec:intro},  the conventional definition
of the day-of-the-week effect refers to the abnormal negative or
abnormal positive returns on Monday and Friday, respectively. Since
not all the markets are open on the same days, comparisons among
countries could be difficult. For example, the Israeli market is
open from Sunday through Thursday (\cite{LauterbachUngar}) and the
first day of the week in the Kuwait Stock Exchange is Saturday
(\cite{Al-Loughani}). Additionally, markets are not open
simultaneously, due to the different time zones (\cite{KohWong}).
Consequently, spill-over effects can influence returns and could
distort results if such influence is not incorporated into the
model.

In order to overcome  these difficulties, we develop here a more
general definition of the day-of-the-week effect that exploits the
potential of the symbolic analysis of time series. Instead of
estimate the return on each day by means of equation \ref{eq:dow},
we will look at the relative position of the return on each day
within its week. If there is no seasonal effect, the order in which
the days appear in each position (from the worst until the best
return of the week), should be random. Otherwise, a seasonal pattern
would be detected.

First, we need to give an specific definition of our seasonal effect.
It must be emphasized  that, according to our proposal, we are not
interested in detecting abnormal negative or positive returns on a
given day. Instead, we are looking for the features of the return on
a given day within its week, from the worst return of the week to
the best return of it, independently of its sign. Thus, a new
definition of the day-of-the-week effect is required.

\begin{def0} \label{def:dow}  The day-of-the-week effect occurs
 whenever a pattern appears much more or much less frequently
  than expected by the uniform distribution. \end{def0}

From this definition a natural null hypotheses arises:

\begin{hypo} \label{hypo:1}  \begin{equation} H_0:\; \sharp(P_1) = \sharp(P_2) = \dots =
\sharp(P_{120})\end{equation} \end{hypo} where $\sharp(P_k)$,
$k=\{1,\dots,120\}$, stands for ``absolute frequency of pattern
$k$".

Since we are interested in studying the day-of-the week effect,
testing this hypothesis is insufficient for our purposes.

We should count the number of times in  which a given  day exhibits
the worst return of the week, the next to worst return, and so on,
until the best return of the week is detected. In other words, we
should count the number of times a given day $i$ occupies the first,
second, third, fourth or fifth position in a pattern and place the
absolute frequencies in a matrix as follows:

\begin{def0}\label{def:matrix} Let $A=\left(a_{ij}\right)$ be a 5x5 matrix.
Element $a_{ij}$ is the absolute frequency of return on day $i$ at the position $j$. \end{def0}

Displaying results in this way, we count how many times a given day
is in position 0 (the worst return of the week), position 1,
position 2, position 3, and position 4 (the best return of the
week). As a consequence, we advance two additional hypothesis:

\begin{hypo} \label{hypo:2}  \begin{equation} H_0:\; a_{i0} = a_{i1} = a_{i2} = a_{i3} = a_{i4}, \; i=0\dots 4 \end{equation} \end{hypo}
This hypothesis says that a given day $i$ could occupy any position, from the worst to the best return within a week.

\begin{hypo} \label{hypo:3} \begin{equation} H_0:\; a_{0j} = a_{1j} = a_{2j} = a_{3j} = a_{4j} \; j=0\dots 4\end{equation} \end{hypo}
This hypothesis says that a given position in the week $j$ could be occupied by any day of the week.

All these null hypotheses could be tested using Pearson's
chi-squared test.  This test is useful to verify if there is a
significant difference between an expected frequency distribution
and an observed frequency distribution. Following \cite{Emma} the
test statistic is:
\begin{equation}
Q_{\{j,i\}}=\sum_{\{i,j\}=0}^4\left[\frac{\left(f_{o,ij}-f_{e}\right)^2}{f_{e}} \right]
\label{eq:chi}
\end{equation}
where $f_{o,ij}$ is the observed frequency of day $i$ at position $j$ and $f_{e}$ is the expected frequency $\sum_{k=1}^{120} \sharp(P_k) /5$. $Q$ is distributed asymptotically as a $\chi^2$ with 4 degrees of freedom.

We  advance two additional hypotheses focused on the so-called
``Monday effect''.

\begin{hypo} \label{hypo:4} \begin{equation} \begin{split} H_0:\;  & \sharp(P_{34})+\sharp(P_{36})+\sharp(P_{40})+\sharp(P_{42})+\sharp(P_{46})+\sharp(P_{48})+\\
&\sharp(P_{58})+\sharp(P_{60})+ \sharp(P_{64})+\sharp(P_{66})+\sharp(P_{70})+\sharp(P_{72})+\\
&\sharp(P_{82})+\sharp(P_{84})+\sharp(P_{88})+\sharp(P_{90})+ \sharp(P_{94})+\sharp(P_{96})+\\
&\sharp(P_{106})+\sharp(P_{108})+\sharp(P_{112})+\sharp(P_{114})+\sharp(P_{118})+\sharp(P_{120})
= (24/120) N \end{split} \end{equation} \end{hypo} This hypothesis
tests whether patterns with Monday having the largest return are
preferred patterns or not. Pattern numbers $P_{xx}$ correspond to those displayed in Table \ref{tab:ordinalpatterns}.

\begin{hypo} \label{hypo:5} \begin{equation}
 H_0:\; \sharp(P_1)+\sharp(P_3)+\sharp(P_7)+\sharp(P_9)+\sharp(P_{13})+\sharp(P_{15})=(6/120) N \end{equation}
  \end{hypo}
This hypothesis tests whether the patterns with Monday exhibiting
the lowest weekly return and Friday the  largest are preferred
patterns or not.

These hypotheses are tested using the binomial test,  which for
large samples can be approximated by the normal distribution. The
test statistic  is:
\begin{equation}
z=\frac{p_e-p_o}{\sqrt{\frac{p_o q_o}{N}}}
\label{eq:binomial}
\end{equation}
where $p_o$ is the observed frequency, $q_o=1-p_o$, $p_e$ is the
expected frequency,   and $N$ is the number of ``weeks'', i.e. the
number of 5-day patterns in the sample.

 Our definition assumes that the day-of-the-week effect could be produced by the dependence among days of the same week. However, by splitting time series into weeks, we implicitly assume the independence among weeks. Even though this later assumption may be questionable, we do this way in order to emphasize the order of the returns within the week. We could relax this assumption, by moving data daily, instead of weekly. In this way, we could compare if, e.g. Friday in week t influences Monday in week t+1 However, it could result in a more confuse analysis, and we let it for further research.

\begin{table}[htbp]
  \centering
  \caption{Ordinal patterns. Each number $\{0, 1, 2, 3, 4\}$ of a pattern represents a day of the week, beginning on Monday. The position of the numbers in a pattern represents the increasing order of returns within a week.}
    \begin{tabular}{rl|rl|rl|rl|rl}
    \toprule
    \multicolumn{1}{l}{$P_{xx}$} & Pattern & \multicolumn{1}{l}{$P_{xx}$} & Pattern & \multicolumn{1}{l}{$P_{xx}$} & $P_{xx}$ & \multicolumn{1}{l}{$P_{xx}$} & Pattern & \multicolumn{1}{l}{$P_{xx}$} & Pattern \\
    \midrule
    1     & 01234 & 25    & 10234 & 49    & 20134 & 73    & 30124 & 97    & 40123 \\
    2     & 01243 & 26    & 10243 & 50    & 20143 & 74    & 30142 & 98    & 40132 \\
    3     & 01324 & 27    & 10324 & 51    & 20314 & 75    & 30214 & 99    & 40213 \\
    4     & 01342 & 28    & 10342 & 52    & 20341 & 76    & 30241 & 100   & 40231 \\
    5     & 01423 & 29    & 10423 & 53    & 20413 & 77    & 30412 & 101   & 40312 \\
    6     & 01432 & 30    & 10432 & 54    & 20431 & 78    & 30421 & 102   & 40321 \\
    7     & 02134 & 31    & 12034 & 55    & 21034 & 79    & 31024 & 103   & 41023 \\
    8     & 02143 & 32    & 12043 & 56    & 21043 & 80    & 31042 & 104   & 41032 \\
    9     & 02314 & 33    & 12304 & 57    & 21304 & 81    & 31204 & 105   & 41203 \\
    10    & 02341 & 34    & 12340 & 58    & 21340 & 82    & 31240 & 106   & 41230 \\
    11    & 02413 & 35    & 12403 & 59    & 21403 & 83    & 31402 & 107   & 41302 \\
    12    & 02431 & 36    & 12430 & 60    & 21430 & 84    & 31420 & 108   & 41320 \\
    13    & 03124 & 37    & 13024 & 61    & 23014 & 85    & 32014 & 109   & 42013 \\
    14    & 03142 & 38    & 13042 & 62    & 23041 & 86    & 32041 & 110   & 42031 \\
    15    & 03214 & 39    & 13204 & 63    & 23104 & 87    & 32104 & 111   & 42103 \\
    16    & 03241 & 40    & 13240 & 64    & 23140 & 88    & 32140 & 112   & 42130 \\
    17    & 03412 & 41    & 13402 & 65    & 23401 & 89    & 32401 & 113   & 42301 \\
    18    & 03421 & 42    & 13420 & 66    & 23410 & 90    & 32410 & 114   & 42310 \\
    19    & 04123 & 43    & 14023 & 67    & 24013 & 91    & 34012 & 115   & 43012 \\
    20    & 04132 & 44    & 14032 & 68    & 24031 & 92    & 34021 & 116   & 43021 \\
    21    & 04213 & 45    & 14203 & 69    & 24103 & 93    & 34102 & 117   & 43102 \\
    22    & 04231 & 46    & 14230 & 70    & 24130 & 94    & 34120 & 118   & 43120 \\
    23    & 04312 & 47    & 14302 & 71    & 24301 & 95    & 34201 & 119   & 43201 \\
    24    & 04321 & 48    & 14320 & 72    & 24310 & 96    & 34210 & 120   & 43210 \\
    \bottomrule
    \end{tabular}%
  \label{tab:ordinalpatterns}%
\end{table}%

\section{Simulation of fractional Brownian motion \label{sec:simulation}}

In this section we apply the above-outlined technique to simulated
time series. We used MATLAB$^\copyright$ {\ttfamily wfbm} function,
in order to simulate fractional Brownian motion for ${\mathcal
H}=\{0.1,\dots,0.9\}$, where $\mathcal H$ is the Hurst exponent. Then, we take first differences in the time
series in order to obtain the corresponding fractional Gaussian
noise (fGn). The Hurst exponent $H$ characterizes the scaling behavior of the range of cumulative departures of a time series from its mean. The study  of long range dependence can be traced back to seminal paper by \citet{Hurst51}, whose original  methodology was applied to detect long memory in hydrologic time series. This method was also explored by \cite{Mandel68} and later introduced in the study of economic time series by \cite{Mandel72}. If the series of first differences is a white noise, then its ${\mathcal H}= 0.5$. Alternatively, Hurst exponents greater than 0.5 reflect persistent processes and less than 0.5 define antipersistent processes.
 
We perform  1000 simulations consisting of 10000
data-points for each value of $\mathcal{ H}$. Accordingly, we obtain
2000 ``weeks'', which will be classified into one of the $5!=120$
possible patterns. If the underlying stochastic process is purely
random and uncorrelated, the frequency of patterns should be
uniform. On the contrary, if some correlation is present, some
patterns could be preferred over others. All the tests are performed
at a $5\%$ significance level.

In Table \ref{tab:hypothesis1}  we test the equality of patterns
(Hypothesis \ref{hypo:1}). When ${\mathcal H}=0.5$ (ordinary
Gaussian noise), we cannot reject, on average, the null hypothesis
of equal appearance of patterns. Out of the 1000 simulations, only
in 50 cases is the null hypothesis rejected. When we move away form
${\mathcal H}=0.5$, in both directions, rejection increases almost
symmetrically. This clearly shows that some kind of correlation
affects the distribution of ordinal patterns.

As commented in the previous section, this analysis is not
sufficient.  Therefore, we proceed to test Hypothesis \ref{hypo:2}
and present the pertinent results in Table \ref{tab:hypothesis2}. We
test the hypothesis (for every ${\mathcal H}$ value) for each of the
samples and for the average of the samples. We find that for
${\mathcal H}=0.5$ we cannot reject the null hypothesis. In fact,
rejection occurs in only 51 to 59 times out of 1000 samples. In
other words, when the generating stochastic process is a white
noise, any day is equally prone to occupy any of the positions in
the pattern. This is the same to say that any day could exhibit the
best or the worst return of the week, or any intermediate value
among them. When we move away from the value 0.5, rejections
increase. However, the effect is stronger for ${\mathcal H}>0.5$
than for ${\mathcal H}<0.5$. This could mean that a positive long
range correlation (i.e. a persistent time series) is more likely to
exhibit a day-of-the-week behavior than anti-persistent time series.
Additionally, Monday, Wednesday and Friday are the days most
affected by the value-change of  $\mathcal{ H}$.

Regarding Hypothesis \ref{hypo:3}, results are displayed in Table
\ref{tab:hypothesis3}.  In the case of the uncorrelated process
(${\mathcal H}=0.5$), one encounters that how good or bad is the
return within a trading week is independent of the day of the week.
This hypothesis is only rejected only 59 times out of the 1000
simulations. When we move away, and then correlations become
stronger, patterns exhibits some degree of preference, increasing
the number of rejections. As in the case of Hypothesis
\ref{tab:hypothesis2}, rejections are more frequent in case of
persistent time series.

Hypothesis \ref{hypo:4} tests whether the presence of patterns with
Monday as the largest return is in agreement with the uniform
distribution. There are 24 patterns with Monday as the last element.
Consequently, its expected frequency is 0.2. Table
\ref{tab:Molargest} displays the results of the simulations. In the
case of a pure Gaussian noise, we cannot reject the null hypothesis,
as in only 117 out of the 1000 trials we reject it. However,
increasing the Hurst exponent produces an increment  in the number
of rejections. Additionally, we observe that larger Hurst values are
associated with greater observed frequency of patterns with Monday
as largest return. However, we cannot reject the null hypothesis
until $H=0.8$ or higher.

Hypothesis  \ref{hypo:5} tests  the presence of a weekly seasonality
with Monday as the smallest return of the week and Friday as the
largest. There are 6 patterns with this structure. In analogy with
the preceding finding, for an uncorrelated noise, this pattern is
neither a preferred nor a rare one. Nevertheless, the increase of
the Hurst exponent produces a quick increase in the number of
rejections: 309 out of 1000 when $H=0.7$. More impressive is how
preferred this pattern is in most of the simulations. For $H=0.6$,
in 698 simulations, the observed frequency of these 6 patters was
above expectation, and for $H=0.7$, $p_o>p_e$ in 873 simulations.

Symbolic analysis is powerful  to detect nontrivial hidden
correlations in data. As shown by \cite{RossoCarpiB,Carpi20102020} a
correlated structure as produced by fractional Gaussian noise
processes generates an uneven presence of patterns. Provided a
sufficiently long time series, no pattern is forbidden. However,
strongly correlated structure produces the emergence of preferred
and rare patterns.

Our artificial time series are larger than the usual data-sets used
in economics. Consequently, the presence of preferred patterns as
the ones evaluated in Hypothesis  \ref{hypo:5} casts doubts on the
validity of previous findings of day-of-the-week. In particular, we
claim that, in view of our results, the day-of-the-week effect is
mainly produced by a complex correlation structure of the pertinent
data.

\begin{table}
  \caption{Test of Hypothesis \ref{hypo:1} on simulated series}
\centering
\begin{tabular}{lrllrl}
    \toprule
    $\mathcal{H}$     &       &       & $\mathcal{H}$     &       &  \\
    \midrule
    \multirow{3}[0]{*}{0.1} & $\chi^2$   & 224.82856$^{***}$ & \multirow{3}[0]{*}{0.6} & $\chi^2$   & 20.42699 \\
                   & \#rejections & 1000  &       & \#rejections & 348 \\
    \multirow{3}[0]{*}{0.2} & $\chi^2$   & 140.50932$^{*}$ & \multirow{3}[0]{*}{0.7} & $\chi^2$   & 86.57005 \\
                            & \#rejections & 1000  &       & \#rejections & 996 \\
    \multirow{3}[0]{*}{0.3} & $\chi^2$   & 67.89548 & \multirow{3}[0]{*}{0.8} & $\chi^2$   & 204.51209$^{***}$ \\
                           & \#rejections & 970   &       & \#rejections & 1000 \\
    \multirow{3}[0]{*}{0.4} & $\chi^2$   & 18.10740 & \multirow{3}[0]{*}{0.9} & $\chi^2$   & 386.14031$^{***}$ \\
                            & \#rejections & 317   &       & \#rejections & 1000 \\
    \multirow{3}[0]{*}{0.5} & $\chi^2$   & 0.13315 &       &       &  \\
                             & \#rejections & 50    &       &       &  \\
    \bottomrule \\
				\multicolumn{4}{l}{$^{*,**,***}$: significant at 10\%, 5\% and 1\% level.}
\end{tabular}%
\label{tab:hypothesis1}%
\end{table}%

\begin{table}
  \caption{Test of Hypothesis \ref{hypo:2} on simulated series}  
  \centering
\begin{tabular}{lrlllll}
    \toprule
    $\mathcal{H}$     & \multicolumn{1}{c}{} & \multicolumn{1}{c}{M} & \multicolumn{1}{c}{T} & \multicolumn{1}{c}{X} & \multicolumn{1}{c}{T} & \multicolumn{1}{c}{F} \\
    \midrule
    \multirow{3}[0]{*}{0.1} & $\chi^2$   & 5.47646 & 1.94578 & 5.40877 & 2.06613 & 5.80607 \\
          & \#rejections & 461   & 176   & 437   & 176   & 464 \\
    \multirow{3}[0]{*}{0.2} & $\chi^2$   & 3.83407 & 1.34267 & 3.30037 & 1.33138 & 3.83153 \\
          & \#rejections & 308   & 124   & 275   & 139   & 318 \\
    \multirow{3}[0]{*}{0.3} & $\chi^2$   & 1.95798 & 0.62585 & 1.79598 & 0.59399 & 2.03637 \\
          & \#rejections & 186   & 87    & 146   & 86    & 162 \\
    \multirow{3}[0]{*}{0.4} & $\chi^2$   & 0.62419 & 0.17731 & 0.47961 & 0.22011 & 0.60417 \\
          & \#rejections & 82    & 54    & 78    & 57    & 71 \\
    \multirow{3}[0]{*}{0.5} & $\chi^2$   & 0.00666 & 0.00304 & 0.00078 & 0.00500 & 0.00159 \\
          & \#rejections & 51    & 53    & 54    & 57    & 59 \\
    \multirow{3}[0]{*}{0.6} & $\chi^2$   & 0.76829 & 0.25922 & 0.67148 & 0.25509 & 0.87231 \\
          & \#rejections & 95    & 68    & 79    & 66    & 93 \\
    \multirow{3}[0]{*}{0.7} & $\chi^2$   & 3.94460 & 1.16245 & 3.40845 & 1.22150 & 3.72159 \\
          & \#rejections & 318   & 123   & 293   & 114   & 320 \\
    \multirow{3}[0]{*}{0.8} & $\chi^2$   & 9.90281$^{**}$ & 3.44099 & 8.96147$^{*}$ & 3.61403 & 9.96604$^{**}$ \\
          & \#rejections & 706   & 270   & 657   & 295   & 719 \\
    \multirow{3}[0]{*}{0.9} & $\chi^2$   & 20.42595$^{***}$ & 7.95607$^{*}$ & 20.65903$^{***}$ & 7.86588$^{*}$ & 20.71557$^{***}$ \\
          & \#rejections & 960   & 572   & 966   & 596   & 966 \\
    \bottomrule \\
		\multicolumn{5}{l}{$^{*,**,***}$: significant at 10\%, 5\% and 1\% level.}
    \end{tabular}%
  \label{tab:hypothesis2}%
\end{table}%

\begin{table}
  \caption{Test of Hypothesis \ref{hypo:3} in simulated series}
  \centering
   \begin{tabular}{crlllll}
    \toprule
    $\mathcal{H}$     & \multicolumn{1}{c}{} & \multicolumn{1}{c}{worst return} & \multicolumn{1}{c}{} & \multicolumn{1}{c}{} & \multicolumn{1}{c}{} & \multicolumn{1}{c}{best return} \\
    \midrule
    \multirow{3}[0]{*}{0.1} & $\chi^2$   & 5.66761 & 2.40628 & 4.47306 & 2.35024 & 5.80600 \\
          & \#rejections & 461   & 193   & 357   & 188   & 471 \\
    \multirow{3}[0]{*}{0.2} & $\chi^2$   & 3.69734 & 1.58367 & 2.83837 & 1.46318 & 4.05746 \\
          & \#rejections & 299   & 133   & 220   & 144   & 335 \\
    \multirow{3}[0]{*}{0.3} & $\chi^2$   & 2.08654 & 0.68270 & 1.44706 & 0.80686 & 1.98700 \\
          & \#rejections & 187   & 95    & 117   & 82    & 181 \\
    \multirow{3}[0]{*}{0.4} & $\chi^2$   & 0.48517 & 0.21129 & 0.42396 & 0.22292 & 0.76205 \\
          & \#rejections & 73    & 60    & 68    & 51    & 89 \\
    \multirow{3}[0]{*}{0.5} & $\chi^2$   & 0.00390 & 0.00301 & 0.00335 & 0.00009 & 0.00673 \\
          & \#rejections & 58    & 57    & 59    & 49    & 37 \\
    \multirow{3}[0]{*}{0.6} & $\chi^2$   & 0.78880 & 0.34282 & 0.58457 & 0.24124 & 0.86898 \\
          & \#rejections & 95    & 66    & 85    & 56    & 91 \\
    \multirow{3}[0]{*}{0.7} & $\chi^2$   & 3.72085 & 1.27291 & 3.09096 & 1.35108 & 4.02279 \\
          & p-value & 0.44510 & 0.86595 & 0.54272 & 0.85265 & 0.40293 \\
          & \#rejections & 294   & 115   & 264   & 127   & 315 \\
    \multirow{3}[0]{*}{0.8} & $\chi^2$   & 10.08089$^{**}$ & 3.67766 & 8.59256$^{*}$ & 3.66076 & 9.87346$^{**}$ \\
          & \#rejections & 700   & 306   & 631   & 296   & 700 \\
    \multirow{3}[0]{*}{0.9} & $\chi^2$   & 20.63329$^{***}$ & 8.33357$^{*}$ & 20.41387$^{***}$ & 7.67672 & 20.56505$^{***}$ \\
          & \#rejections & 965   & 617   & 963   & 570   & 962 \\
    \bottomrule \\
		\multicolumn{5}{l}{$^{*,**,***}$: significant at 10\%, 5\% and 1\% level.}
    \end{tabular}%
  \label{tab:hypothesis3}%
\end{table}%

\begin{table}
\caption{Test of Hypothesis 4 in simulated series}
\centering
\begin{tabular}{lllllll}
    \toprule
    Hurst & $p_e$  & $p_0$  & $q_0$  & $z$     &  \# of rejections & \# $p_o>p_e$ \\
    \midrule
    0.10  & 0.20  & 0.18850 & 0.81150 & 0.97729 &  306   & 161 \\
    0.20  & 0.20  & 0.19034 & 0.80966 & 0.81787 &  232   & 222 \\
    0.30  & 0.20  & 0.19397 & 0.80603 & 0.50693 &  172   & 275 \\
    0.40  & 0.20  & 0.19742 & 0.80258 & 0.21572 &  124   & 359 \\
    0.50  & 0.20  & 0.20243 & 0.79757 & -0.20089 &  117   & 507 \\
    0.60  & 0.20  & 0.20842 & 0.79158 & -0.68858 &  107   & 670 \\
    0.70  & 0.20  & 0.21406 & 0.78594 & -1.13915 &  225   & 839 \\
    0.80  & 0.20  & 0.22125 & 0.77875 & -1.70120$^{**}$  & 380   & 924 \\
    0.90  & 0.20  & 0.22867 & 0.77133 & -2.26805$^{**}$  & 627   & 983 \\
    \bottomrule \\
						\multicolumn{5}{l}{$^{*,**,***}$: significant at 10\%, 5\% and 1\% level.}
    \end{tabular}%
  \label{tab:Molargest}%
\end{table}%

\begin{table}
\caption{Test of Hypothesis 5 in simulated series} 
  \centering
\begin{tabular}{lllllll}
    \toprule
    Hurst & $p_e$  & $p_0$  & $q_0$  & $z $     & \# of rejections & \# $p_o>p_e$ \\
    \midrule
    0.1   & 0.05  & 0.04013 & 0.95987 & 1.67154$^{**}$  & 562   & 45 \\
    0.2   & 0.05  & 0.04263 & 0.95737 & 1.21287  & 427   & 91 \\
    0.3   & 0.05  & 0.04531 & 0.95469 & 0.74944  & 302   & 155 \\
    0.4   & 0.05  & 0.04824 & 0.95176 & 0.27363  & 174   & 290 \\
    0.5   & 0.05  & 0.05197 & 0.94803 & -0.29438  & 78    & 507 \\
    0.6   & 0.05  & 0.05588 & 0.94412 & -0.85028  & 119   & 698 \\
    0.7   & 0.05  & 0.06074 & 0.93926 & -1.49392$^{*}$  & 309   & 873 \\
    0.8   & 0.05  & 0.06681 & 0.93319 & -2.23719$^{**}$  & 589   & 983 \\
    0.9   & 0.05  & 0.07376 & 0.92624 & -3.01987$^{***}$  & 856   & 998 \\
    \bottomrule \\
						\multicolumn{5}{l}{$^{*,**,***}$: significant at 10\%, 5\% and 1\% level.}
    \end{tabular}%
  \label{tab:MoxxxFr}%
\end{table}%


\section{Empirical application \label{sec:empirical}}

We use daily data of NYSE Composite Price Index from 03/01/1966 to
08/12/2017,  for a total of 13,550 observations. All data used in
this paper was retrieved from Data-Stream. We split the sample into
four non-overlapping periods of equal length (3,050 data points), and a last period of 1350 datapoints, in
order to verify the temporal evolution of the seasonal effect. We
compute daily log returns in order to apply our test.

\begin{figure}
\begin{center}
\includegraphics[width=0.95\textwidth]{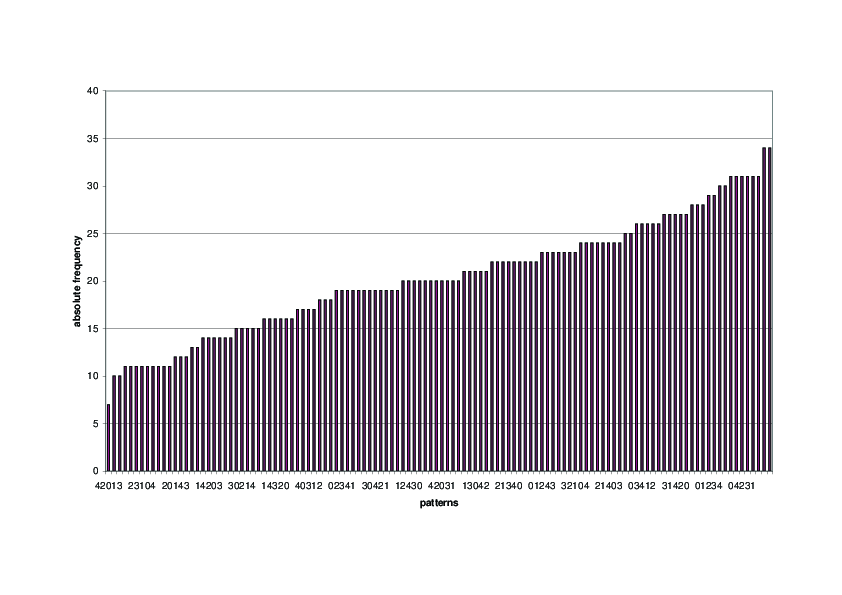}
\caption{Histogram of pattern frequency for the whole period.}
\label{fig:histogram}
\end{center}
\end{figure}

Regarding Hypothesis \ref{hypo:1} (see Table \ref{tab:allpatterns}),
we find, in the whole sample, no forbidden patterns. Under these
circumstances, we should discard chaotic behavior in the time series
(\cite{RossoCarpiB}) The least frequent pattern, with an absolute
frequency equal to 7, is 42013 (i.e.
$r_{Fr}<r_{We}<r_{Mo}<r_{Tu}<r_{Th}$). The most frequent patterns,
with an absolute frequency equal to 34, are 03421 and 04312 (i.e.
$r_{Mo}<r_{Th}<r_{Fr}<r_{We}<r_{Tu}$ and
$r_{Mo}<r_{Fr}<r_{Th}<r_{Tu}<r_{We}$, respectively).  As stated in
Section \ref{sec:op}, if data were generated at random, i.e., if no
seasonal effect exists, patterns should uniformly appear,
configurating the histogram of a uniform distribution. However, as
seen in Figure \ref{fig:histogram}, ours is a far from uniform
distribution.

Table \ref{tab:total} exhibits frequencies  and tests for hypothesis
\ref{hypo:2} and hypothesis \ref{hypo:3}. Following the horizontal
lines of the table, we test whether a given day indifferently
occupies any position in the returns of the week. Along  the
vertical sense of the table, we test whether a given position within
a week is indifferently occupied by any day.

Regarding the whole period we observe  that we cannot accept the
null hypothesis of equal distribution of returns across the week. In
fact, if we observe Table \ref{tab:total}, Monday acquires the
lowest return of the week more frequently than any other week-day.

In order to justify the fact that intrinsic temporal  correlations
play a significant role in the ordinal patterns, we have also
estimated the frequency of the patterns for the shuffled return
data. ``Shuffled" realizations of the original data are obtained by
permuting them in  random order, and eliminating, consequently, all
non-trivial temporal correlations. From Table \ref{tab:shuffling},
we observe that patterns are distributed in a more or less uniform
fashion and, consequently, we cannot reject the null hypotheses.
Therefore, the results of our test are not due to  chance.

If we analyze the evolution of the daily seasonal  behavior through
time, it is clear that the day-of-the-week effect  disappears in
daily returns of the NYSE Composite index. Results are reflected in
Tables \ref{tab:subperiod1}, \ref{tab:subperiod2},
\ref{tab:subperiod3}, \ref{tab:subperiod4} and \ref{tab:subperiod5}. Considering the last
subperiod,  only  Tuesday effect remains. Tuesday is the
most frequent day in the worst position and Friday tends to occupy
the best return within each week. However, we cannot reject
that the worst return of the week can be occupied by any other week-
day. According to this analysis we observe, in agreement with the
literature, a disappearing weekly effect in daily returns in the US
market. This disappearing effect is related to the hidden underlying
dynamics of data, rather than with markets participants behavior, as
it was classically envisaged in the literature. We would like to
emphasize that our test unveils the hidden correlation structure of
daily returns. As in the case of the artificial generated series,
the pattern behavior in real time series is strongly affected by the
long memory of data. 

An important difference between real and simulated data is that, whereas in the controlled experiment the Hurst exponent is, by definition, constant across all the time series, in the case of real data, the Hurst exponent tends to vary across time (see e.g. \cite{CajuTab,CajuTabEM,BaGuMa2012}). This situation makes difficult the direct comparison between both results. Moreover, we can observe that the power of the test is more sensitive for ${\mathcal H}>0.5$ in detecting the Monday effect (see Table \ref{tab:MoxxxFr} ). In fact, for ${\mathcal H}=0.9$, the test rejects 856 out of the 1000 simulated series.  Another factor that influences results is the time series length. As recalled by \cite{RossoCarpiB}, short time series could result in the incorrect detection of forbidden patterns.

\begin{table}
  \caption{Absolute frequency of each pattern. Whole period.  Each number $\{0, 1, 2, 3, 4\}$ of a pattern represents a day of the week, beginning on Monday. The position of the numbers in a pattern represents the increasing order of returns within a week.}
 \centering
\begin{tabular}{rr|rr|rr|rr}
    \toprule
    Pattern & Abs. Freq. & Pattern & Abs. Freq. & Pattern & Abs. Freq. & Pattern & Abs. Freq. \\
    \midrule
    42013 & 7     & 14320 & 16    & 42031 & 20    & 21403 & 24 \\
    23041 & 10    & 20314 & 16    & 42301 & 20    & 32140 & 24 \\
    24013 & 10    & 23140 & 16    & 43012 & 20    & 42310 & 24 \\
    02413 & 11    & 31204 & 16    & 43210 & 20    & 02431 & 25 \\
    13240 & 11    & 04213 & 17    & 10324 & 21    & 04123 & 25 \\
    13402 & 11    & 20413 & 17    & 12043 & 21    & 01423 & 26 \\
    23104 & 11    & 40312 & 17    & 13042 & 21    & 03412 & 26 \\
    30124 & 11    & 41230 & 17    & 14032 & 21    & 20134 & 26 \\
    40123 & 11    & 20431 & 18    & 34201 & 21    & 34012 & 26 \\
    40132 & 11    & 23410 & 18    & 02134 & 22    & 34210 & 26 \\
    41203 & 11    & 40231 & 18    & 03124 & 22    & 01342 & 27 \\
    43102 & 11    & 02143 & 19    & 10432 & 22    & 12403 & 27 \\
    20143 & 12    & 02341 & 19    & 21340 & 22    & 31420 & 27 \\
    24310 & 12    & 03142 & 19    & 21430 & 22    & 34021 & 27 \\
    42103 & 12    & 10342 & 19    & 24301 & 22    & 43201 & 27 \\
    20341 & 13    & 12340 & 19    & 32014 & 22    & 02314 & 28 \\
    23014 & 13    & 23401 & 19    & 34102 & 22    & 10423 & 28 \\
    13024 & 14    & 24031 & 19    & 41302 & 22    & 21304 & 28 \\
    14203 & 14    & 30421 & 19    & 01243 & 23    & 01234 & 29 \\
    24103 & 14    & 32041 & 19    & 03241 & 23    & 34120 & 29 \\
    24130 & 14    & 41032 & 19    & 10234 & 23    & 10243 & 30 \\
    31024 & 14    & 42130 & 19    & 12034 & 23    & 14023 & 30 \\
    41320 & 14    & 43120 & 19    & 14302 & 23    & 01432 & 31 \\
    30142 & 15    & 01324 & 20    & 21034 & 23    & 04132 & 31 \\
    30214 & 15    & 12430 & 20    & 32104 & 23    & 04231 & 31 \\
    30412 & 15    & 31042 & 20    & 03214 & 24    & 30241 & 31 \\
    40213 & 15    & 31402 & 20    & 04321 & 24    & 31240 & 31 \\
    41023 & 15    & 32401 & 20    & 13204 & 24    & 43021 & 31 \\
    12304 & 16    & 32410 & 20    & 14230 & 24    & 03421 & 34 \\
    13420 & 16    & 40321 & 20    & 21043 & 24    & 04312 & 34 \\
    \bottomrule
    \end{tabular}%
    \label{tab:allpatterns}%
\end{table}%

\begin{table}
 \caption{Absolute frequency of each day in each position. Whole period: 03/01/1966--08/12/2017. Columns 2 to 6 reflect the frequency a given day is the worst return of its week, the next to worst return, etc. until the best return of its week. $Q$ is the $\chi^2$ statistic defined in equation \ref{eq:chi}. Hurst=0.5471.}
  \centering
\begin{tabular}{lllllll}
    \toprule
& \multicolumn{2}{l}{worst return}       &     &     \multicolumn{2}{r}{best return}&  $Q $    \\
      \midrule
   Mo    & 637   & 503   & 537   & 501   & 532   & 22.78967$^{***}$ \\
    Tu   & 557   & 572   & 475   & 509   & 597   & 17.94834$^{***}$ \\
    We    & 479   & 540   & 564   & 564   & 563   & 9.92989$^{**}$ \\
    Th    & 570   & 505   & 564   & 581   & 490   & 12.66052$^{**}$ \\
    Fr    & 467   & 590   & 570   & 555   & 528   & 16.74908$^{***}$ \\
    Total & 2710  & 2710  & 2710  & 2710  & 2710  &  \\
    $Q$     & 36.21402$^{***}$ & 11.25092$^{**}$ & 11.56089$^{**}$ & 9.12177$^{*}$ & 11.92989$^{**}$ &  \\
    \bottomrule \\
\multicolumn{4}{l}{$^{*,**,***}$: significant at 10\%, 5\% and 1\% level.}
    \end{tabular}%
  \label{tab:total}%
\end{table}%

\begin{table}
  \caption{Absolute frequency of each day in each position with shuffled data for the whole period. Columns 2 to 6 reflect the frequency a given day is the worst return of its week, the next to worst return, etc. until the best return of its week. $Q$ is the $\chi^2$ statistic defined in equation \ref{eq:chi}.}
  \centering
\begin{tabular}{lllllll}
    \toprule
& \multicolumn{2}{l}{worst return}       &     &     \multicolumn{2}{r}{best return}&  $Q $    \\
      \midrule
    Mo  & 506 & 469 & 501 & 484 & 480 & 1.91393  \\
    Tu & 488 & 508 & 472 & 498 & 474 & 1.95082  \\
    We  & 513 & 475 & 491 & 471 & 490 & 2.24590  \\
    Th  & 479 & 491 & 509 & 482 & 479 & 1.32787  \\
    Fr  & 454 & 497 & 467 & 505 & 517 & 5.75410  \\
    Total & 2440 & 2440 & 2440 & 2440 & 2440 &       \\
    $Q $ & 4.47951 & 2.09016 & 2.69672 & 1.49590 & 2.43033 &       \\
    \bottomrule \\
\multicolumn{4}{l}{$^{*,**,***}$: significant at 10\%, 5\% and 1\% level.}
    \end{tabular}%
  \label{tab:shuffling}%
\end{table}%

\begin{table}
  \caption{Absolute frequency of each day in each position. Subperiod 1: 03/01/1966--09/09/1977. Columns 2 to 6 reflect the frequency a given day is the worst return of its week, the next to worst return, etc. until the best return of its week. $Q$ is the $\chi^2$ statistic defined in equation \ref{eq:chi}. Hurst=0.5633}
  \centering
\begin{tabular}{lllllll}
    \toprule
& \multicolumn{2}{l}{worst return}       &     &     \multicolumn{2}{r}{best return}&  $Q$     \\
    \midrule
    Mo  & 182 & 121 & 105 & 97  & 105 & 39.37705$^{***}$  \\
    Tu & 108 & 145 & 107 & 124 & 126 & 7.95082$^{*}$  \\
    We  & 108 & 99  & 120 & 142 & 141 & 12.21311$^{**}$  \\
    Th  & 116 & 119 & 138 & 132 & 105 & 5.65574  \\
    Fr  & 96  & 126 & 140 & 115 & 133 & 9.72131$^{**}$  \\
    Total & 610 & 610 & 610 & 610 & 610 &       \\
   $ Q$  & 38.55738$^{***}$ & 8.88525$^{*}$ & 9.00000$^{*}$ & 9.65574$^{**}$ & 8.81967$^{*}$ &       \\
    \bottomrule \\
\multicolumn{4}{l}{$^{*,**,***}$: significant at 10\%, 5\% and 1\% level.}
    \end{tabular}%
  \label{tab:subperiod1}%
\end{table}%

\begin{table}
\caption{Absolute frequency of each day in each position. Subperiod 2: 12/09/1977--19/05/1989.Columns 2 to 6 reflect the frequency a given day is the worst return of its week, the next to worst return, etc. until the best return of its week. $Q$ is the $\chi^2$ statistic defined in equation \ref{eq:chi}. Hurst=0.5168} 
  \centering
\begin{tabular}{lllllll}
    \toprule
& \multicolumn{2}{l}{worst return}       &     &     \multicolumn{2}{r}{best return}&  $Q$     \\
    \midrule
    Mo  & 165 & 111 & 110 & 107 & 117 & 19.37705$^{***}$  \\
    Tu & 138 & 117 & 111 & 104 & 140 & 8.60656$^{*}$  \\
    We  & 100 & 125 & 129 & 131 & 125 & 5.18033  \\
    Th  & 112 & 119 & 129 & 148 & 102 & 10.11475$^{**}$  \\
    Fr  & 95  & 138 & 131 & 120 & 126 & 8.90164$^{*}$  \\
    Total & 610 & 610 & 610 & 610 & 610 &       \\
  $  Q $  & 28.01639$^{***}$ & 3.44262 & 3.63934 & 10.73770$^{**}$ & 6.34426 &       \\
    \bottomrule \\
\multicolumn{4}{l}{$^{*,**,***}$: significant at 10\%, 5\% and 1\% level.}
    \end{tabular}%
  \label{tab:subperiod2}%
\end{table}%

\begin{table}
\caption{Absolute frequency of each day in each position. Subperiod 3: 22/05/1989--26/01/2001. Columns 2 to 6 reflect the frequency a given day is the worst return of its week, the next to worst return, etc. until the best return of its week. $Q$ is the $\chi^2$ statistic defined in equation \ref{eq:chi}. Hurst=0.4453}
  \centering
 \begin{tabular}{lllllll}
    \toprule
  & \multicolumn{2}{l}{worst return}       &     &     \multicolumn{2}{r}{best return}&  $Q$   \\
      \midrule
    Mo  & 109 & 105 & 133 & 126 & 137 & 6.72131  \\
    Tu & 136 & 124 & 103 & 119 & 128 & 4.96721  \\
    We  & 89  & 141 & 147 & 125 & 108 & 18.68852$^{***}$  \\
    Th  & 154 & 117 & 108 & 123 & 108 & 11.81967$^{**}$  \\
    Fr  & 122 & 123 & 119 & 117 & 129 & 0.68852  \\
    Total & 610 & 610 & 610 & 610 & 610 &       \\
   $ Q $  & 20.31148$^{***}$ & 5.57377 & 10.75410$^{**}$ & 0.49180 & 5.75410 &       \\
    \bottomrule \\
	\multicolumn{4}{l}{$^{*,**,***}$: significant at 10\%, 5\% and 1\% level.}
    \end{tabular}%
  \label{tab:subperiod3}%
\end{table}%

\begin{table}
  \caption{Absolute frequency of each day in each position. Subperiod 4: 29/01/2001--05/10/2012. Columns 2 to 6 reflect the frequency a given day is the worst return of its week, the next to worst return, etc. until the best return of its week. $Q$ is the $\chi^2$ statistic defined in equation \ref{eq:chi}. Hurst= 0.4983} 
 \centering
   \begin{tabular}{lllllll}
    \toprule
& \multicolumn{2}{l}{worst return}       &     &     \multicolumn{2}{r}{best return}&  $Q$    \\
    \midrule
    Mo  & 134 & 106 & 121 & 128 & 121 & 3.59016  \\
    Tu & 112 & 139 & 117 & 106 & 136 & 7.09836  \\
    We  & 126 & 115 & 125 & 115 & 129 & 1.40984  \\
    Th  & 131 & 105 & 121 & 125 & 128 & 3.40984  \\
    Fr  & 107 & 145 & 126 & 136 & 96  & 13.45902$^{***}$  \\
    Total & 610 & 610 & 610 & 610 & 610 &       \\
   $ Q $  & 4.63934 & 11.57377$^{**}$ & 0.42623 & 4.47541 & 7.85246$^{*}$ &       \\
    \bottomrule \\
						\multicolumn{4}{l}{$^{*,**,***}$: significant at 10\%, 5\% and 1\% level.}

    \end{tabular}%
  \label{tab:subperiod4}%
\end{table}%

\begin{table}[htbp]
  \centering
  \caption{Absolute frequency of each day in each position. Subperiod 5: 08/10/2012--08/12/2017. Columns 2 to 6 reflect the frequency a given day is the worst return of its week, the next to worst return, etc. until the best return of its week. $Q$ is the $\chi^2$ statistic defined in equation \ref{eq:chi}. Hurst= 0.4914}
    \begin{tabular}{lllllll}
		    \toprule
          & \multicolumn{1}{l}{worst return} &       &       &       & \multicolumn{1}{l}{best return} & \multicolumn{1}{l}{$Q$} \\
    \midrule
    Mo    & 47    & 60    & 68    & 43    & 52    & 7.51852 \\
    Tu.   & 63    & 47    & 37    & 56    & 67    & 10.96296$^{**}$ \\
    We    & 56    & 60    & 43    & 51    & 60    & 3.81481 \\
    Th    & 57    & 45    & 68    & 53    & 47    & 6.22222 \\
    Fr    & 47    & 58    & 54    & 67    & 44    & 6.18519 \\
    Total & 270   & 270   & 270   & 270   & 270   &  \\
    $Q$     & 3.55556 & 4.03704 & 14.85185$^{***}$ & 5.62963 & 6.62963 &  \\
		    \bottomrule \\
						\multicolumn{4}{l}{$^{*,**,***}$: significant at 10\%, 5\% and 1\% level.}
    \end{tabular}%
  \label{tab:subperiod5}%
\end{table}%

In the Supplementary Material file  we present the simulation and
test of hypotheses for shorter time series. Additionally, we perform
an exhaustive analysis of 83 stock indices with different Hurst
levels. We can observe that greater Hurst-levels are associated with
more significant presence of preferred patterns.

It is clear that theoretical and empirical analyses exhibit some differences. We have to acknowledge that real stock markets dynamics do no follow a pure fGn. In fact, there is not only long range dependence in financial time series, but also in volatility, as shown recently by \cite{Bariviera2017ELett}. Precisely, more advanced models such as the fractional normal tempered stable process presented by \cite{Kim2012AML,Kim2015Frontiers}, allow for long range dependence in both volatility and noise, and asymmetric dependence structure for the joint distribution. There are many economic variables that influence behaviors known as ``stylized facts'' of financial time series: volatility clustering, fat tails, asymmetric dependence, etc. For example, \cite{Kim2016AMF} finds that long range dependence increases more in volatile markets, during the Lehman Brothers collapse.

We try to emphasize in this paper that, even using a simple model such as a fGn,  some part of the seasonal effect is simply due  to the correlation structure of data, and not only by economic reasons. This finding could be used as a starting point in further research, in order to apply prewhitening to time series previous to its analysis, in order to obtain more reliable results.

\section{Conclusions \label{sec:conclusion}}

We propose a more general definition of the day-of-the-week effect.
We use symbolic time series analysis in order to develop a test to
detect it. According to Definition \ref{def:dow}, this effect takes
place when a pattern is much more or much less frequent than
expected from the uniform distribution. The nature of the seasonal
effect is reflected in a frequency matrix (Definition
\ref{def:matrix}), and a $\chi^2$ test is performed. The new
definition allows for a more general and comprehensive study on
return seasonality. 
We would like to highlight that the methodology we use here is unique in that it is nonlinear, ordinal, requires no {\it a priori} model. Additionally, it provides statistical results in term of a probability density function. To the extent of our knowledge, no one using time series analysis has used a similar methodology before.

Both, theoretical and empirical applications show that this method could be useful to discriminate between rare and
preferred patterns of a time series.  We show that the so-called day-of-the-week effect is influenced not only by traders' behavior or economic variables. It could be also be induced by the stochastic generating process of data. The findings in this paper could be taken into account in future research, aiming at the separation between the economics causes and the long-range correlation causes of this financial phenomenon.

\bibliographystyle{apalike}
\bibliography{dayoftheweek}

\end{document}